\documentclass[aps,prb,twocolumn,showpacs,superscriptaddress]{revtex4}
\usepackage{graphicx}%

\begin{document}

\title{Magneto-elastic polarons in the hole-doped quasi-1D model system Y$_{2-x}$Ca$_x$BaNiO$_5$}

\author{F.-X. Lannuzel}
\affiliation{ Institut des Mat\'{e}riaux Jean Rouxel, 2 rue de la Houssiniere, 44322 Nantes, France}
\author{E. Janod}
\affiliation{ Institut des Mat\'{e}riaux Jean Rouxel, 2 rue de la Houssiniere, 44322 Nantes, France}
\author{C. Payen} 
\affiliation{ Institut des Mat\'{e}riaux Jean Rouxel, 2 rue de la Houssiniere, 44322 Nantes, France}
\author{B. Corraze}
\affiliation{ Institut des Mat\'{e}riaux Jean Rouxel, 2 rue de la Houssiniere, 44322 Nantes, France}
\author{D. Braithwaite}
\affiliation{ CEA Grenoble, DRFMC/SPSMS, 38054 Grenoble, France}
\author{O. Chauvet}
\affiliation{ Institut des Mat\'{e}riaux Jean Rouxel, 2 rue de la Houssiniere, 44322 Nantes, France}
\date{\today}

\begin{abstract}
Charge transport in the hole-doped quasi-1D model system Y$_{2-x}$Ca$_x$BaNiO$_5$ (x $\leq$ 0.15) is investigated in the 50-300 K temperature range. The resistivity temperature dependence is characterized by a constant activation energy $E_{a}/k_{B}\sim$ 1830 K at room temperature while $E_{a}$ decreases upon cooling. We suggest that $E_{a}$ measures the binding energy of the doped holes which form magneto-acoustic polarons when polarizing the neighboring Ni spins. A semi-classical model is proposed which allows to relate the electrical measurements and the bulk magnetic susceptibility. This model gives a picture of the spin-charge-lattice relation in this inhomogeneously doped quasi-1D system and explains its unusual one-particle charge excitation spectrum close to the Fermi level.
\end{abstract}


\maketitle
%
%    INTRODUCTION
%
\section{introduction}
\label{introduction}

Recent developments in the field of strongly correlated electron systems provide strong indications that a broad class of doped transition-metal-oxides exhibits intrinsically inhomogeneous electronic states \cite{dagotto, tranquada}. Strong interactions in these materials drive phase separation between regions of distinct electronic structure or the formation of nanoscale inhomogeneities in the form of, e.g., stripes. The study of inhomogeneous systems and the identification of spin-charge-lattice relations are difficult tasks, and it is important to establish such relations in model materials. Due to their relative simplicity, doped one-dimensional antiferromagnets are good candidates for such a purpose.

Among these systems, Y$_{2}$BaNiO$_{5}$ is an attractive case. The crystal structure of this quasi-1D Haldane antiferromagnet is built from chains of compressed apex-sharing NiO$_{6}$ octahedra running along the z-axis \cite{amador}. Despite a significant Ni$(3d_{3z^2-r^2})$-O$(2p_z)$ orbital interaction \cite{mattheiss}, this compound is a charge transfer insulator with a gap\cite{maiti} of 2 eV due to strong on-site Coulomb repulsions. The Ni$^{2+}$ (3d$^{8}$) $S=1$ spins are antiferromagnetically coupled along the chains through Ni-O-Ni superexchange with $J_{AF}/k_{B}\sim$ 300 K, leading to a spin-liquid ground state with a Haldane spin-gap \cite{darriet, ditusa} $\Delta_{H}$=100 K. Hole doping in Y$_{2}$BaNiO$_{5}$, which is achieved through off-chain substitution of Ca$^{2+}$ (or Sr$^{2+}$) for Y$^{3+}$, has motivated many experimental and theoretical works. Dc conductivity measurements \cite{ditusa} have shown that the doped hole remains localized. However, no charge ordering occurs in this transition metal oxide\cite{xu, ito}. Optical conductivity measurements have shown an absence of conductivity at zero-energy in the optical spectrum while a charge gap of the order of 0.3 eV survives 
the doping \cite{ito}. Actually, X-ray absorption spectroscopy (XAS)\cite{ditusa, lannuzel, hu}, photoemission\cite{maiti, fagot} as well as band structure calculations \cite{novac} have demonstrated that doping generates new empty states with a mixed Ni$(3d_{3z^2-r^2})$-O$(2p_z)$ character in the charge gap. Hole doping of these antibonding states results in a shortening of the mean Ni-O distances along the chain direction ($z$-axis)\cite{lannuzel, alonso}. The presence of localized holes has also a severe influence on the spin background. It induces a strong ferromagnetic interaction between the two Ni spins on both sides of the hole \cite{xu}, and new magnetic states appear within the spin gap \cite{ditusa, xu}. These new magnetic states contribute to a low-temperature sub-Curie behavior of the bulk magnetic susceptibility which depends on the doping level \cite{janod}. The overall picture which emerges includes thus the presence of a hole which locally disturbs the spin chain by inducing a localized spin polarization (presumably a lattice distortion as well) while the magnetic and electronic states of the other parts of the chain are kept essentially unchanged at low doping level \cite{xu, ito}. Indeed transport of such a localized charge should be influenced by the spin background. The purpose of this paper is to address this problem.

Here we report dc conductivity measurements of Y$_{2-x}$Ca$_x$BaNiO$_5$ single crystals and ceramic samples with $0\leq x \leq 0.15$. The ceramic samples were already studied by static magnetic susceptibility measurements \cite{janod}. We argue that conduction is due to hopping of holes associated with their polarization cloud and lattice distortion, \textit{i.e.}  magneto-elastic polarons. This picture is consistent with the spectroscopic investigations which show an absence of spectral weight at the Fermi level. We propose a simple model which allows to relate the transport properties to the static magnetic susceptibility of the antiferromagnetic medium. At low doping, a characteristic size of the polarization cloud of the order of 3 Ni-Ni distance is extracted from the model while a local ferromagnetic coupling of the order of 1500 K is found. 
%
%    EXPERIMENTAL DETAILS
%
\section{experimental details}
\label{experimental details}

Small single crystals of Y$_{2-x}$Ca$_x$BaNiO$_{5+\delta}$ (x=0, 0.07 and 0.15) are grown using a conventional flux method \cite{yokoo}, where the crystals are furnace cooled at the end of the growth. The undoped crystal ($x$=0) is then annealed at 800$^\circ$C in air and quenched to room temperature. The Ca content is determined by wavelength and energy dispersive x-ray analysis. Four-probe contacts are made using silver-charged epoxy resin annealed at 400$^\circ$C during one hour. The resulting off-stoichiometry $\delta$ is found to be smaller than 0.012 (0.006) in the doped (undoped) crystals. The resistivity is measured in the parallel 
(z axis) or perpendicular to the chains geometry with a home made guarded set-up which allows resistance measurements up to 10$^{13} \Omega$ from room temperature down to $\sim$40 K (measurements at lower temperatures were prohibited by high impedances and by very long time constants). The results were reproduced on several crystals and they are in good agreement with published data \cite{ditusa, ito} above 90 K. Note that our resistivity measurement set-up allowed to extend the temperature range of the measurements down to $\sim$ 40 K, \textit{i.e.} below the spin gap temperature.  Measurements under quasi-hydrostatic pressure were performed in a Bridgman type cell with tungsten carbide anvils and steatite as the pressure transmitting medium. Ceramic samples with $x$ in the range 0.05-0.15 were prepared and characterized as described in Ref.\onlinecite{lannuzel}. The electrical resistances of rectangular shaped sintered pellets were measured by using the four-probe method described above.

\begin{figure}
\centering
\includegraphics[width=8cm]{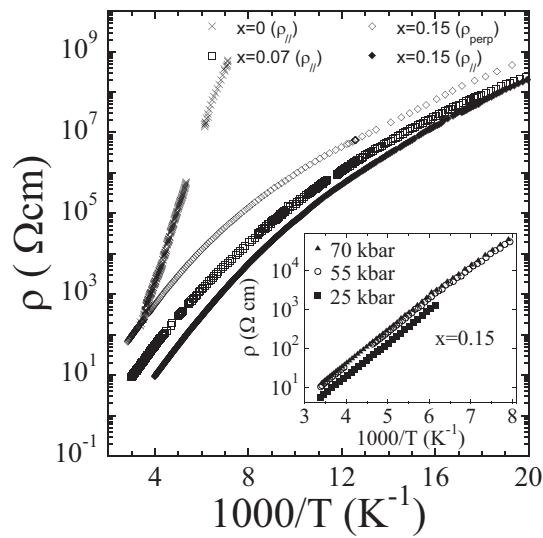}
\caption{Temperature dependence of the resistivity of Y$_{2-x}$Ca$_x$BaNiO$_5$ single-crystals ($x=0, 0.07$ and $0.015$) in the direction parallel $\rho_{\|}$ or 
perpendicular $\rho_{\perp}$ to the chains. Inset: Pressure dependence of $\rho_{\|}$ for the Y$_{1.85}$Ca$_{0.15}$BaNiO$_5$ sample.}
\label{fig:1}
\end{figure}
%
%    Results and discussion
%
\section{results and discussion}
\label{results and discussion}
\subsection{Transport properties}

As shown in Fig.1, the resistivity along the chain axis $\rho_{\|}$ of the undoped sample follows an activated temperature dependence, $\ln(\rho_{\|}) \propto E_{g}/2k_{B}T$ with $E_{g} \approx$ 0.7 eV. This energy is much lower than the charge transfer gap ($\approx$ 2 eV \cite{maiti}) but it is comparable to $E_{g}\approx$ 0.6 eV found in Ref. \onlinecite{ditusa}. Indeed this behavior is attributed to the weak oxygen off-stoichiometry which persists in the sample as in many transition metal oxides despite the annealing treatment. Actually $E_{g}$ strongly depends on the way the $x=0$ sample was treated. The data reported here correspond to our most insulating crystal. 

Doping does not result in a metallic state as shown in Fig.1 for $x$=0.07 or $x$=0.15. It is in agreement with the localized nature of the doped holes already mentioned. A 1D character of the electrical transport is inferred from the resistivity anisotropy ratio $\rho_{\|}/\rho_{perp}\sim 60$ at 300 K which reflects the 1D chain structure of the material. We tried to apply a quasi hydrostatic pressure on the sample in order to recover a possible metallic state as in the spin ladder Sr$_{0.4}$Ca$_{13.6}$Cu$_{24}$O$_{41}$ where superconductivity appears upon pressure\cite{uehara}. However as shown in inset of Fig.1 for the $x$=0.15 single crystal, neither metallicity nor superconductivity is found when applying  pressure up to 70 kbar. Instead, the pressure application does not significantly affect the resistivity of the sample. 

At ambient pressure, the resistivity of the Ca-doped compounds is lowered by several orders of magnitude at low temperature as compared with the undoped sample resistivity. Indeed it is associated with a new temperature dependence of the resistivity.
\begin{figure}
\centering
\includegraphics[width=8cm]{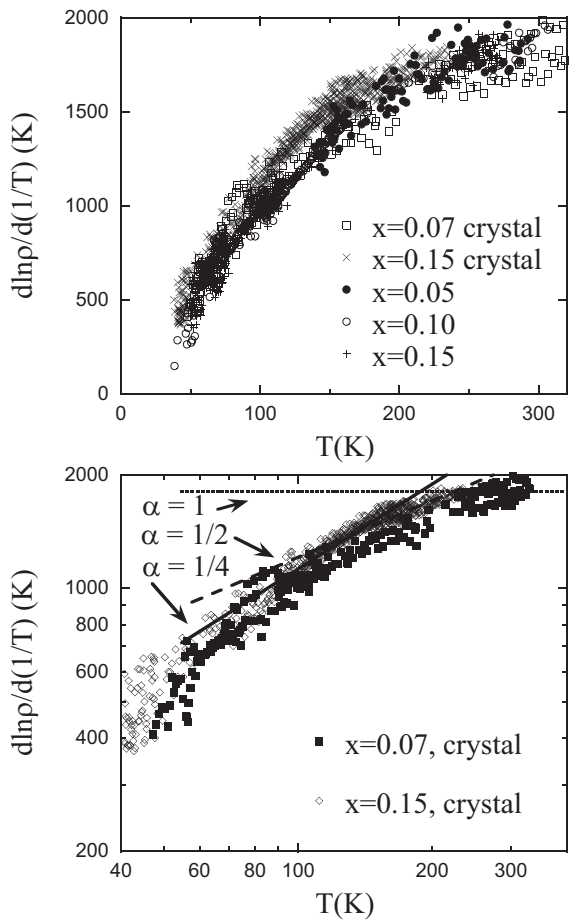}
\caption{Upper panel: Effective activation energy $E_{a}(T)$ versus $T$ for the $x$=0.07 and $x$=0.15 single crystal samples and for the $x$=0.5, 0.10 and 0.15 ceramic samples. Bottom panel: $E_{a}(T)$ versus $T$ for the single crystal samples in log-log scales. The straight lines represent $ln(\rho)\propto (T_{0}/T)^{\alpha}$ variations.}
\label{fig:2}
\end{figure}
It is quite interesting to note that this temperature dependence is the same for all our doped samples whether they are single crystals or ceramic samples. This is shown in upper panel of Fig.2 where the effective activation energy $E_{a}(T)=dln(\rho)/d(1/T)$ is plotted versus temperature for the single crystal ($x$=0.07 and $x$=0.15) and the ceramic samples ($x$=0.5, 0.10 and 0.15). Within the experimental uncertainties, all the dependencies fall on the same curve. The same trend is observed in the literature \cite{ditusa, ito} where the temperature dependence of the resistivity is described within the quasi-1D Mott's variable range hopping (VRH) model.  VRH yields $\rho=\rho_{0}\exp(T_{0}/T)^{\alpha}$ with $\alpha$=1/2 in 1D. A similar functional dependence can also describe 
our data in the same temperature range \textit{i.e.} between 100 K and 250 K. Still there are several reasons for which we believe that VRH is not adequate to describe our results.
Bottom panel of Fig.2 shows $E_{a}(T)$  versus $T$ in logarithmic scales for the single crystal samples. In this representation, a VRH dependence appears as a linear part whose slope gives (1-$\alpha$). As shown in the figure, while VRH with $\alpha\sim$1/2 can be inferred between 100 K and 250 K in agreement with Ref.\onlinecite{ditusa, ito}, extending the temperature range of measurements to lower temperatures requires that $\alpha$ decreases continuously below 100 K while a simple activated behavior ($\alpha$=1) is observed above $\sim$250 K. Indeed attributing this complex temperature dependence to 3D VRH ($\alpha$=1/4) below 100 K followed by 1D VRH ($\alpha$=1/2) up to 250 K and then by simple activation with $E_{a}/k_{B}\simeq$ 1800 K close to room temperature is highly questionable. This plot rather suggests that transport does not occur through VRH at all. Let us remind that VRH is a low temperature process which is valid only below a fraction of the Debye temperature. This is obviously not the case in the present temperature range. It also requires a constant and finite density of states at the Fermi level for which there is no spectroscopic evidence in these compounds. 

Alternatively Fig.2 may suggest that hole transport occurs with a constant activation energy $E_{a}$ which corresponds to a gap $E=2\times E_{a}\simeq$ 0.34 eV close to the charge gap observed in Ref.\onlinecite{ito} above 250 K while the activation energy is reduced below 250 K. We argue here that the doped holes form magneto-elastic polarons. In this temperature range, transport of such dressed charges should occur through multiphonon hopping which gives an activated dependence with  $E_{a}$ given by half of the binding energy. For temperatures smaller than the antiferromagnetic (AF) superexchange constant of the undoped system $T_{AF}=J_{AF}/k_{B}\sim$285 K, short range AF correlations develop which reduces the binding energy hence $E_{a}$. Indeed, this picture is supposed to give the same binding energy for the polarons whatever the doping level as long as the physics of these species is kept constant. It is in agreement with the unique temperature dependence found here in our samples whatever the doping level.
\subsection{Semi-classical model}
Hereafter, we derive a simple 1D model in order to describe the main physics of the charge transport. As suggested by theoretical and experimental works \cite{batista, xu}, we consider that each doped hole introduces a strong local ferromagnetic (FM) interaction $J_{1}$ within the antiferromagnetic chain. We also assume that the FM interaction polarizes the surrounding Ni$^{2+}$ moments over a distance $R$ along the chain. The drastic reduction of the resistivity observed experimentally upon doping suggests that the holes are not confined to single, isolated sites but delocalized over several lattice spacings. $J_{1}$ has to be larger than $J_{AF}$ in order the polarization to occur. Following Mott \cite{Mott}, the energy of the polaron is thus given by:
\begin{eqnarray}
E=\frac{\hbar^{2}\pi^{2}}{2mR^2}-\frac{V_{e}}{R}-J_{1}+W_{2}\frac{R}{a}
\label{eq2}
\end{eqnarray}
where the first term is the kinetic energy, the second term is the elastic energy gain in a continuous medium approximation, the third one is the FM magnetic energy gain, the last one reflects the cost of polarizing $R/a$ moments in the AF chain ($W_{2}$ is thus the AF exchange energy at 0 K, $a$ being the Ni-Ni distance along the chain). The elastic term accounts for the shortening of the Ni-O distance within the polaron, in agreement with the observed shortening of the $a$ cell parameter with hole doping\cite{alonso}. In absence of magnetic interactions ($J_{1}=J_{AF}=0$), the size $R_{0}$ of the (purely acoustic) polaron is obtained by minimizing $E$ with respect to $R$. It gives a binding energy $E_{0}=\hbar^{2}\pi^{2}/(2mR_{0}^2)$. The size of the magneto-elastic polaron is obtained in the same way. In the limit of small differences, $R_{1}\simeq R_{0}(2\beta+2)/(3\beta+2)$ is found where $\beta=(W_{2}R_{0}/a)/E_{0}$ measures the strength of the AF cost normalized to the binding energy of the acoustic polaron. The binding energy (if positive) is thus given by:
\begin{eqnarray}
\frac{E_{1}}{E_{0}}= -\frac{1}{y^{2}}+\frac{2}{y}+\frac{J_{1}}{E_{0}}-\beta y
\label{eq2}
\end{eqnarray}
where $y=R_{1}/R_{0}$ is of the order of unity. Eq.\ref{eq2} is valid if $E_{1}>0$. Indeed, it shows that the magneto-elastic polaron is more stable than the purely acoustic one if the FM energy gain $J_{1}$ overcomes the AF energy cost $W_{2}R/a$.

Let us now discuss the situation at finite temperature. As soon as the temperature is higher than the spin gap of the undoped compound, the AF energy cost is reduced and the binding energy increases with increasing temperature. This is true as long as $T < T_{AF}$. When $T > T_{AF}$ (but for $T < J_{1}/k_{B}$), one should expect that the binding energy remains constant. According to this argument, Fig.2 reflects the temperature dependence of the polaron binding energy $E_1$ ($E_a$ = 1/2 $E_1$), with a constant binding energy above $T_{AF}$. Below $T_{AF}$, the binding energy decreases with the temperature due to increasing AF energy cost. If correct, it means that the transport properties should be related to the magnetic ones. 

\begin{figure}
\centering
\includegraphics[width=8cm]{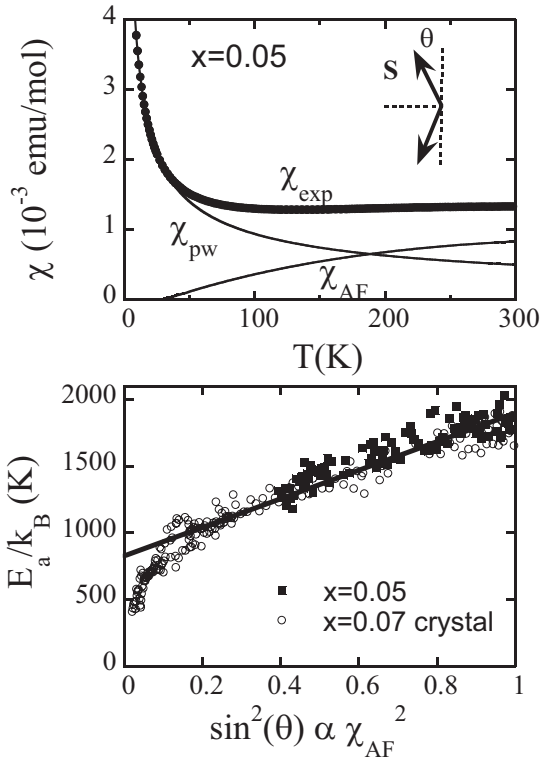}
\caption{Upper panel: Temperature dependence of the magnetic susceptibility $\chi_{exp}$ of the $x$=0.05 ceramic sample. Data taken from ref.\onlinecite{janod}. The solid lines represent the polarons ($\chi_{pw}$) and the AF spin background ($\chi_{AF}$) contributions. 
Bottom panel: Relationship of the effective activation energy of the $x$=0.05 ceramic (black squares) and $x$=0.07 crystal (white dots) samples with the magnetic susceptibility of the AF spin background of the $x$=0.05 ceramic sample}
\label{fig:3}
\end{figure}
\subsection{Relation between transport and magnetic properties}
It was not possible to measure the magnetic susceptibility of the single crystal samples studied here because of their size. The magnetic susceptibilities of the ceramic samples with 0 $\leq x \leq$ 0.20 are presented and discussed in details in Ref.\onlinecite{janod}. Briefly summarized, the undoped compound susceptibility is characteristic of the AF $S$=1 Haldane chain above 30 K, with a thermally activated regime $\chi_{H} \propto T^{-1/2}$exp$(-\Delta/T)$ and a broad maximum close to 380 K. Below 30 K, a Curie contribution attributed to a slight off-stoichiometry shows up. While the susceptibility is only slightly affected by hole doping for $T>T_{AF}$, a strong enhancement of the susceptibility is observed at lower temperatures that masks the thermally activated regime. Obviously this enhancement results from the introduction of holes onto the chain and it is related to the new magnetic states which appear inside the Haldane gap\cite{xu}. 

The upper panel of Fig.3 shows the temperature dependence of the static magnetic susceptibility $\chi_{exp}$ of the $x$=0.05 Ca doped sample. Below $T\sim$ 100 K, $\chi_{exp}$ diverges when the temperature decreases. It has been shown\cite{janod} that $\chi_{exp}$ follows a sub-Curie paramagnetic behavior $\chi_{pw}\propto T^{-\gamma}$ with $\gamma \sim$ 0.6 below 30 K.  This sub-Curie behavior is the signature of low-energy magnetic excitations that exist within and between polarons through the $S$=1 background and it reflects the contribution of the polarons with their polarization cloud. The remaining part of the susceptibility $\chi_{AF}=\chi_{exp}-\chi_{pw}$ is due to the contribution of the AF sea. Below $\sim$200 K, this contribution can be described by the Haldane $\chi_{H}$ term. This indicates that the polarons contribution is disconnected from the AF sea contribution at low doping level. Above 300 K \textit{ i.e.} for $T> T_{AF}$, the compound recovers the expected mean-field paramagnetic behavior.  The three $\chi_{exp}$, $\chi_{pw}$ and  $\chi_{AF}$ contributions are shown in Fig.3 for the $x$=0.05 sample. 

It is important to note that is not possible to separate the AF sea and the polaron contributions for higher doping rates ($x \geq 0.10$). In this case, even if  a sub-Curie behavior can still be identified in a limited temperature range at low temperatures \cite{janod}, $\chi_{exp}-\chi_{pw}$ does not follow the $\chi_{H}$ temperature dependence. It suggests that the polarons contribution cannot be disconnected from the AF sea contribution. It is likely that the isolated polaron model is inadequate due to significant interactions among polarons and possible overlaps of their polarization clouds which may give additional contributions to $\chi_{exp}-\chi_{pw}$.

Let us now try to relate the magnetic properties and the transport properties. Using a classical spins picture in the temperature range of investigation (50 K$<T<$300 K), we can phenomenologically describe the increase of $\chi_{AF}$ as being due to a progressive departure of an AF order (see inset of upper panel of Fig.3). We introduce a phenomenological angle $\theta$ such that $\chi_{AF}\propto sin(\theta)$ with $\theta$ =0 in the singlet ground state while $\theta$ increases as $T$ increases (but below $T_{AF}$). The same approximation allows to quantify the AF energy cost of Eq.\ref{eq2} which is given by: 
\begin{eqnarray}
W_{2}(T)=J_{AF} S_{i}.S_{i+1}\sim W_{2}(0).cos(2\theta)
\label{eq3}
\end{eqnarray}
$W_{2}(T)$ should be zero at $T_{AF}$  where $\chi_{AF}$ is normalized. Here, $sin(\theta)$ is obtained by normalizing $\chi_{AF}$ at room temperature. Eq.\ref{eq3} shows that $W_{2}(T)\propto\chi_{AF}^{2}$. It is thus possible to relate the resistivity activation energy to the magnetic susceptibility through: $E_{a}=A + B \chi_{AF}^{2}$. The bottom panel of Fig.3 presents the activation energy  versus  $sin^{2}(\theta)\propto\chi_{AF}^{2}$ for the $x$=0.05 ceramic sample (black squares). Because of the limited temperature range for the determination of $E_{a}$ for this sample, we add the activation energy data of the $x$=0.07 single crystal sample (white dots). This comparison of the two data sets is justified since $E_{a}$ is found to be rather insensitive to the doping level in the $0.05\leq x \leq 0.15$ range (see Fig.2 and part III.A). A linear relationship between $E_{a}$ and $sin^{2}(\theta)$ is clearly obeyed at least above $\sim$ 80 K. It is shown as a solid line on the figure. 

Physical informations can be extracted from the $A$ and $B$ coefficients. They are found to be 850$\pm$50 K and 950$\pm$50 K respectively.  Within this model, $k_{B}B=W_{2}(0)R_{1}/a$ and $2k_{B}A=(-1/y^{2}+2/y)E_{0}+J_{1}- k_{B}B$. Assuming that $W_{2}(0)/k_{B}=J_{AF}/k_{B}\sim $ 300 K gives the size of the polaron $R_{1}\sim 3.1\pm 0.1 a$ where $a$ is the Ni-Ni distance along the chain. From $R_{1}$ and with $y$=2/3 (limiting value), the elastic energy gain of the acoustic polaron is $E_{0}/k_{B}\sim$ 1500 K. We can thus deduced from $A$ the value of the ferromagnetic interaction $J_{1}/k_{B}\sim 1550 \pm 150$ K. It can be compared to the theoretical results of Ref.\onlinecite{batista} which gives a ferromagnetic interaction between the spin of a Zhang-Rice doublet and its nearest neighbor Ni spin of the order of 3000 K.
\subsection{Discussion} 
Of course, the model derived through equations (1) to (3) is clearly oversimplified. It is based on a semi classical estimate of the energy. It is also a pure 1D model which neglects all the interchain interactions as well as the polaron-polaron interactions. The magnetic moments are considered in a classical way. Since we restrict ourselves to the 50-300 K temperature range and to the lowest doping level investigated here, we believe that these approximations are quite reasonable. Another issue is the microscopic nature of the polarization cloud surrounding the doped hole. At low temperature when the carriers are completely frozen, inelastic neutron scattering (INS) experiments of Xu \textit{et al.}\cite{xu}  provide evidence for an incommensurate double-peaked structure factor for the low-energy magnetic states created by hole doping. The incommensurability can not be ascribed to inhomogeneous spin and charge ordering in the form of stripes as observed in 2D underdoped cuprates and nickelates\cite{tranquada}, for example.  Xu \textit{et al.} \cite{xu} argue that the holes doped into the chain are located in oxygen orbitals, and they induce an effective ferromagnetic interaction between the Ni spins on both sides of the hole. Spin incommensurability arises even without hole order because of the antiferromagnetic polarization clouds that develop around the holes, with the sizes of the AF droplets controlled by the correlation length of the Haldane chain ($\sim 6 a$ at 0 K). Within this static low-temperature picture, the lattice is not coupled to the spin and charge degrees of freedom and the AF cost of the polaron corresponds to the creation of staggered magnetization within a spin-liquid state. 

The approach followed in this paper was to define a fully ferromagnetic polarization cloud associated with an elastic deformation (magneto-acoustic polaron) extending over a variable distance $R$ into the chain. This allowed to account for the magnetic energy cost due to hole doping in a simple way. Anyhow, the existence of magneto-elastic polarons may explain the unusual low-energy spectral features in the one-particle charge excitation spectrum of Y$_{2-x}$Ca$_x$BaNiO$_5$. In contrast with other doped 
nickelates and cuprates, no finite spectral weight appears in optical conductivity \cite{ito} and photoemission \cite{maiti, fagot} measurements at $E_F$. In these experiments excitation of the polaronic charge releases the polarization energy which is not recovered in the excited state (or at least not in the same way). Hence, there is no low energy spectral weight despite a finite density of states at the Fermi level. Actually, a similar conclusion was drawn by Ito \textit{et al.} \cite{ito} who suggests that the low energy gap in the optical conductivity arises from the bound nature of the holes. 

\section{Summary}
\label{summary}
In summary, transport properties of the hole doped Haldane chain Y$_{2-x}$Ca$_x$BaNiO$_5$ have been investigated down to low temperatures. We argue that transport is due to hopping of magneto-elastic polaronic holes. We propose a simple and semi-classical model which allows to relate the transport properties to the magnetic ones and which explains the unusual spectroscopic features close to the Fermi level. It gives a simple picture of the spin-charge-lattice relation associated with inhomogeneous charge doping in a quasi-1D model system.

\end{document}